\documentclass[11pt,prl,showpacs,nofootinbib,a4paper]{revtex4}
\usepackage{amsmath}
\usepackage{amsfonts}
\usepackage{amssymb}
\usepackage{graphicx} 
\usepackage{times}
\begin{document}
%
%
\date{September 11th 2006} 
\title{THE CABIBBO ANGLE : AN ALGEBRAIC CALCULATION}
\author{Q. Duret}
\email[E-mail: ]{duret@lpthe.jussieu.fr}
\author{B. Machet}
\email[E-mail: ]{machet@lpthe.jussieu.fr}
\affiliation{Laboratoire de Physique Th\'eorique et Hautes \'Energies,
(Paris, France)}
\altaffiliation[UMR 7589 (CNRS / Universit\'e Pierre
et Marie Curie-Paris6 / Universit\'e
Denis Diderot-Paris7).\\ Postal address: ]{LPTHE tour 24-25, 5\raise 3pt
\hbox{\tiny \`eme} \'etage,
          Universit\'e P. et M. Curie, BP 126, 4 place Jussieu,
          F-75252 Paris Cedex 05 (France)} 
\begin{abstract}
We show that the Cabibbo angle $\theta_c$ satisfies the
relation $\tan(2\theta_c)=\pm1/2$ when
universality for diagonal neutral currents of mass eigenstates is satisfied
at the same level of accuracy as the absence of their off-diagonal
counterparts. The predicted value is $\cos\theta_c \approx 0.9732$, only
$7/10000$ away from experimental results. No mass ratio  appears in the
calculation. $\theta_c$ occurs {\em a priori}
for both quark species, and, showing that one recovers 
the standard dependence of leptonic and semi-leptonic decays of
 pseudoscalar mesons, we advocate, like for neutrinos and charged leptons,
for a symmetrical treatment of $u$ and $d$-type quarks.
\end{abstract}
\maketitle
%

\section{Introduction}

Why mixing angles are what they are is, in addition to their more and more
precise experimental determination, one of the most active present
domain of research, both in the leptonic and hadronic sectors.
The attempts that have been 
proposed up to now in order to address this fundamental question, 
such as the linking of the sine of the Cabibbo angle \cite{Cabibbo}
 to the $d$ and $s$ quark mass ratio \cite{Weinberg}, 
are often based on empirical evidence, and so remain unsatisfying
\footnote{The rigorous treatment \cite{MachetPetcov} shows that only
weaker ``asymptotic'' relations hold, which involve both $m_d/m_s$ and
$m_u/m_c$.}
. 
Indeed, there are actually too many free parameters in the most
 general Yukawa couplings of the standard model to infer mixing angles 
from mass ratios.

In \cite{DuretMachet1}, we have shown that maximal mixing angles naturally
arise for non-degenerate coupled systems of physical particles with two
generations, like neutrinos, when one requires universality for the diagonal
 neutral currents of {\em mass eigenstates} and the absence of their
off-diagonal counterparts. This follows from an important feature that we
also demonstrated there, {\it i.e.} that the mixing matrix of such systems 
should never be parametrized as unitary. Such a property
had also been checked before \cite{MaNoVi} for neutral kaons.

In this letter, we would like to suggest one promising 
 step on the path to a better 
understanding of the enigma raised above. 
In direct continuation of \cite{DuretMachet1}, still working in the simple case of two generations,
 we investigate non-degenerate coupled
systems that do not exist on-shell, like quarks. Those systems, as shown in 
\cite{DuretMachet1}, 
are more naturally endowed with a unitary mixing matrix, due to the fact that 
their two mixing angles finally satisfy $\theta_2 = \pm \theta_1 [\pi]$.
Staying in an infinitesimal, but still 2-dimensional,
 neighborhood of such a ``Cabibbo-case'' (as referred to in 
\cite{DuretMachet1}), we impose 
the physical requirement that universality for weak diagonal neutral
currents of mass eigenstates is realized with the same accuracy as the
absence of the off-diagonal mass changing neutral currents (MCNC's).
As a consequence the value of the unique mixing angle is shown to be
very close to that of the measured Cabibbo angle.

\section{Generalities, states and (weak) currents}

In quantum field theory, the states of definite mass for a system of particles 
are defined as the eigenstates, at its poles, of the full (renormalized) 
propagator, which is given by a matrix of dimension $2n_f$ in flavour space.
For non-degenerate coupled systems, like neutral kaons, leptons or quarks,
these mass eigenstates belong to different bases \cite{DuretMachet1}.
Indeed, one finds only one such eigenstate at each of the poles of 
the propagator ; it is, among the $2n_f$ eigenstates of the latter, 
the one corresponding to the vanishing eigenvalue. Hence, the mass eigenstates
 do not make up an orthonormal basis, and
the mixing matrix, which by definition connects the
set of mass eigenstates to that of flavour eigenstates, is non-unitary. 

Nevertheless, for systems of coupled fields which, like quarks,
are never on-shell,  a natural basis appears : the one that occurs at
 any given $z=q^2$. It is then
orthonormal as soon as the inverse propagator $L^{(2)}(z)$ is hermitian.
Thus, the mixing matrix relating it to the flavour basis is unitary 
(accorded that the flavour basis is orthonormal, too). 
This has, in the simple case of two generations, 
the following consequences.
While, in general, ($2 \times 2$) mixing matrices for coupled systems 
are to be parametrized with two mixing angles $\theta_1$ and $\theta_2$,
quark-like (Cabibbo-like) systems finally shrink to one dimension.

As stressed in \cite{DuretMachet1}, the physics of mixing angles is underlain
by weak currents in  mass space. These we shall consider again now.
Neutral (left-handed) weak currents for mass eigenstates are determined by the
combinations $K_1^\dagger K_1$ and $K_2^\dagger K_2$, where $K_1$ and $K_2$
are the mixing matrices for the two types of fermions involved
(for example neutral and charged leptons and, later, for our purpose,
quarks of the $u$ and $d$-type), whereas charged currents give rise to the 
combination matrix $K=K_1^\dagger K_2$.
As soon as neither $K_1$ nor $K_2$ is unitary, the conditions for
universality  (equality of diagonal neutral currents) and absence of
off-diagonal neutral currents, which are built-in properties in flavour space,
 appear on the contrary no longer trivial in the space of mass
eigenstates.

Let us parametrize like in \cite{DuretMachet1} ($c_1$ stands for example
for $\cos\theta_1$)
\begin{equation}
K_1 = \left(\begin{array}{rr} e^{i\alpha}c_1 & e^{i\delta}s_1 \cr
-e^{i\beta}s_2 & e^{i\gamma}c_2
\end{array}\right).
\label{eq:Kparam}
\end{equation}
It is then straightforward to show that universality in mass space for
weak neutral currents imposes the vanishing of
\begin{equation}
R = c_1^2 + s_2^2 - c_2^2 -s_1^2,
\end{equation}
while the absence of MCNC's requires that of
\begin{equation}
S = c_1s_1 -c_2s_2\quad \text{or}\quad T = c_1s_1 + c_2s_2.
\end{equation}
%

\section{Getting the Cabibbo angle from an extra touch of physics}

Let us now demand  that the two conditions for universality and for
the absence of MCNC's  are
realized with the same accuracy. We consider accordingly the equations
\begin{equation}
|R|=|S| \Leftrightarrow R\pm S =0 \quad \text{or}\quad |R|=|T|
\Leftrightarrow R\pm T =0,
\label{eq:cond0}
\end{equation}
{\it i.e.}
\begin{subequations}
\begin{equation}
\cos(2\theta_1) - \cos(2\theta_2) \pm  \frac{1}{2}\big(\sin(2\theta_1) - 
\sin(2\theta_2)\big) = 0,
\label{eq:cond1a}
\end{equation}
\begin{equation}
\cos(2\theta_1) - \cos(2\theta_2) \pm  \frac{1}{2}\big(\sin(2\theta_1) + 
\sin(2\theta_2)\big)=0, 
\label{eq:cond1b}
\end{equation}
\end{subequations}
which directly lead  to the following conditions :
\begin{subequations}
\begin{equation}
\sin(\theta_1 - \theta_2)\big(\cos(\theta_1 + \theta_2) \mp  
2\sin(\theta_1 + \theta_2)\big) = 0,
\label{eq:cond2a}
\end{equation}
\begin{equation}
\sin(\theta_1 + \theta_2)\big(\cos(\theta_1 - \theta_2) \mp  
2\sin(\theta_1 - \theta_2)\big) = 0.
\label{eq:cond2b}  
\end{equation}
\end{subequations}

First, one immediately sees that a trivial solution of the equations
(\ref{eq:cond2a}) and (\ref{eq:cond2b}) is, resp., 
$\theta_2 = \theta_1 \mod \pi$ and 
$\theta_2 = - \theta_1 \mod \pi$. 
This corresponds to Cabibbo-like solutions, for which the two conditions
$R=0$ and $T$ or $S=0$ are separately satisfied, ensuring universality 
for diagonal neutral currents in mass space, and the absence of MCNC's.

Let us now expand (\ref{eq:cond2a}) and (\ref{eq:cond2b}) in the vicinity of 
these Cabibbo-like solutions by writing respectively in there
 $\theta_2 = \pm \theta_1 + \epsilon$ :
\begin{subequations}
\begin{equation}
\label{eq:eps1}
\epsilon \Big( \big( \cos(2\theta_1) \mp 2\sin(2\theta_1) \big) - 
2\epsilon \big( \sin(2\theta_1) \pm 2\cos(2\theta_1) \big) \Big) = 0,
\end{equation}
\begin{equation}
\label{eq:eps2}
\epsilon \Big( \big( \cos(2\theta_1) \mp 2\sin(2\theta_1) \big) + 
2\epsilon \big(\sin(2\theta_1) \pm 2\cos(2\theta_1) \big) \Big) = 0,
\end{equation}
\end{subequations}
and impose that the two conditions (\ref{eq:eps1},\ref{eq:eps2})
are satisfied with the
same accuracy  ${\mathit o}(\epsilon^2)$. This leads  to the following
equation (we replace hereafter $\theta_1$ with $\theta_c$) :
\begin{equation}
\tan(2\theta_c) = \pm \frac{1}{2},
\label{eq:tan2}
\end{equation}
that can be rewritten
\begin{equation}
\big( \tan(\theta_c) \big)^2 \pm 4\tan(\theta_c) - 1 = 0.
\label{eq:tan}
\end{equation}
Eqs.~(\ref{eq:tan}) admit the solutions : $\tan(\theta_c) = \pm(-2 \pm
\sqrt5)$, the smaller of which (they have opposite signs) have the
following value for their cosine : 
$\cos\big(\pm\arctan(2 - \sqrt5)\big) \simeq 0.9732$. 
This result lies remarkably close to the present experimental range
\cite{PDG} 
for the cosine of the Cabibbo angle : $[0.9739, 0.9751]$. The other solution
corresponds to ${\theta_c -\pi}/{2}$, {\em i.e.} to a simple
relabeling of the two corresponding mass states (for example $u_m
\leftrightarrow c_m$).
It is noticeable that this approach yields a constant value for the Cabibbo
angle, which should be confronted with experiment. The most natural, {\em a
priori} $q^2$-dependent, orthonormal basis related to Cabibbo-like systems
mentioned in section 2, should then also exhibit special
properties with respect to its $q^2$ dependence. This will be investigated
in a forthcoming work.

A similar demonstration  holds for $K_2$.

\section{Plea for a symmetrical treatment of both quark species}

We already advocated in \cite{DuretMachet1} for a symmetrical treatment  of
charged and neutral leptons, with maximal mixing for both species. The PMNS
matrix then reduces to the diagonal unit matrix, without putting in
jeopardy the interpretation of the solar neutrino deficit on earth in terms
of neutrino oscillations.

We advocate for the same attitude in the quark sector.
First, there is not a single
good reason to assume that mass and flavour eigenstates coincide for one
species and not for the other one. Secondly, the Cabibbo angle naturally arises
by the mechanism explained above for both quark types.
Last, it is easy to demonstrate that, by a suitable analysis, we recover
standard results. We give below the example of semi-leptonic and leptonic
decays of the $K^+$ meson.

From the most common example of neutral kaons \cite{belusevic},
two  species of mesons are known to occur in physics : flavor and mass
eigenstates.  The first are produced by strong interactions
(which commute with flavour) while  the second also identify with
electroweak eigenstates
\footnote{This is not to be confused with the convention that flavor
states of quarks and leptons are the ones which diagonally couple,
by definition, to weak gauge bosons.}
. When studying its semi-leptonic decays,
the $K^+$ is produced by strong interactions while electroweak 
eigenstates are detected ;
the following subscripts should be accordingly attributed to this reaction :
$K^+_f \rightarrow \pi^0_m \ell^+_m \nu_{em}$.
In a quark-spectator model,  quark flavors are by
definition weakly coupled with $diag(1,1)$ ; the decay channels of $K^+_f$
can only be, accordingly, $(d_f \bar s_f)\, \ell^+ \nu_\ell$ and
$(u_f \bar c_f)\, \ell^+ \nu_\ell$, with coefficient $1$ since
no mixing angle occurs.  When  projected on mass states, these final
flavour states  undergo the Cabibbo rotations which take place for both
quark species.  This gives
 $(d_f \bar s_f) \equiv K^0_f =
c_{\theta_c}^2 K^0_m + c_{\theta_c} s_{\theta_c} \big((s_m \bar s_m) -
(d_m \bar d_m)\big)
-s_{\theta_c}^2 \overline{K^0_m},\  (u_f \bar c_f) \equiv \overline{D^0_f}
 = c_{\theta_c}^2
\overline{D^0_m} + c_{\theta_c} s_{\theta_c}\big((c_m \bar c_m) - (u_m \bar
u_m)\big) - s_{\theta_c}^2 D^0_m$. All semi-leptonic decays of
the $K^+_f$ mesons are then
  described with their observed dependence on the Cabibbo angle (the ones
into $\pi^0$, with their $\cos\theta_c \sin\theta_c$ dependence,
are built up from the $u\bar u$ and $d \bar d$ final states).
We also find  double-Cabibbo-suppressed  $|\Delta S| = 2$ transitions.

Leptonic decays are likewise easily explained by the decomposition
$K^+_f \equiv (u_s \bar s_f) = c_{\theta_c}^2 K^+_m + c_{\theta_c}
s_{\theta_c}(D^+_{s\,m} - \pi^+_m) -s_{\theta_c}^2 D^+_m$ and
with the leptonic decay of the $\pi^+_m$ that takes place with a
coefficient $1$ since, now, no mixing angle occurs in charged weak currents.

This supports our point of view according to which  the Cabibbo
angle appears at the level of both quark species but not in the charged
weak couplings of the  Lagrangian. While the usual ``Cabibbo matrix'', which
connects mass eigenstates for $u$ and $d$-type quarks,
reduces then to the unit matrix, the distinction between flavour and
mass eigenstates for mesonic bound states must be  carefully performed
\footnote{Since the Cabibbo-angle keeps acting between flavour eigenstates
of one type and mass eigenstates of the other type, the evaluation of
``box-diagram'' controlling, for example, the transitions between the
flavour $K^0$ and $\overline{K^0}$ mesons stays unchanged since internal
quark lines are {\em propagating}, that is {\em mass} eigenstates,
while external lines are the $d_f$ and $s_f$ {\em flavour} states.}
.

\section{Conclusion and prospects}

Realizing, as was first done in \cite{MaNoVi}, that mixing matrices of
non-degenerate coupled systems should not be parametrized as unitary, led
in \cite{DuretMachet1} to uncover maximal mixing of leptons as
a class of solutions of very simple physical conditions for their
mass eigenstates.

Now, we have shown that the measured value of the Cabibbo angle $\theta_c$
 is such
that, in its vicinity (allowing the second mixing angle $\theta_2$
to be close to $\theta_1=\theta_c$), universality of diagonal
neutral currents for mass eigenstates is realized with the same  
${\mathit o}\big((\theta_2-\theta_1)^2\big)$ accuracy as the absence of MCNC's. 

In this elementary algebraic calculation, no mass ratio appears.
It may thus help to provide independent information on the latter.
On another side, this feature is  welcome for quark-like systems which
cannot be defined on-shell and for which, accordingly, the notion of
physical mass is ill-defined.

Like for leptons, we  advocate that the two species of quarks
should be treated on equal footing. Indeed, the Cabibbo angle, as a mixing
angle, must occur for both, while it should not appear  inside charged
weak couplings of mass eigenstates.

This work, together with \cite{DuretMachet1}, strongly suggests that the
observed values of mixing angles for quarks and leptons follow from
well defined physical requirements.
The generalization to three generations is currently under investigation.

%
\begin{em}

\end{em}


\begin{thebibliography}{50}
%
\bibitem{Cabibbo}
N. CABIBBO :
Phys. Rev. Lett. 10 (1963) 531-533.

\bibitem{Weinberg}
S. WEINBERG : ``The problem of mass'', in : A Festschrift for I.I. Rabi,
Trans. N.Y. Acad. Sci. (Ser. II) (1977) 185-201.

\bibitem{MachetPetcov}
B. MACHET \& S.T. PETCOV :
 Phys. Lett. B 513 (2001) 371-380, and references therein.

\bibitem{DuretMachet1}
Q. DURET \& B. MACHET : ``Mixing Angles and Non-Degenerate Coupled Systems of
Particles'', hep-ph/0606303, and references therein.

\bibitem{MaNoVi}
B. MACHET, V.A. NOVIKOV \& M.I. VYSOTSKY :
Int. J. Mod. Phys. A 20 (2005) 5399-5452, hep-ph/0407268,
and references therein.

\bibitem{PDG}
PARTICLE DATA GROUP : Review of Particle Physics, Phys. Lett. B 592 (2004)
p.130.

\bibitem{belusevic}
See for example :
R. BELU\u{S}EVI\'C : ``Neutral Kaons'', Springer Tracts in Modern Physics
153 (Springer-Verlag Berlin Heidelberg New-York, 1999), p.8,
 and references therein.


\end{thebibliography}
\end{document}